\documentclass{emulateapj}
\newcommand      \Angstrom       {\,{\rm \AA}}
\newcommand       \mum          {\,{\rm \mu m}}
\newcommand       \rmin         {r_{\rm min}}
\newcommand       \rmax         {r_{\rm max}}
\newcommand       \K            {\,{\rm K}}
\newcommand       \AU           {\,{\rm AU}}
\newcommand       \pc           {\,{\rm pc}}
\newcommand       \g            {\,{\rm g}}
\newcommand       \s            {\,{\rm s}}
\newcommand       \km           {\,{\rm km}}
\newcommand       \cm           {\,{\rm cm}}
\newcommand       \erg          {\,{\rm erg}}

\newcommand       \amin         {a_{\rm min}}
\newcommand       \amax         {a_{\rm max}}
\newcommand       \simali       {\sim\,}

\newcommand       \rp           {r_{\rm p}}
\newcommand       \Mearth       {\,{\rm M_\earth}}
\newcommand       \Mjup         {\,{\rm M_{Jup}}}

\begin{document}

\title{The Dust and Gas Around $\beta$ Pictoris}
\author{C.\ H.\ Chen\altaffilmark{1}, A. Li\altaffilmark{2},
        C. \ Bohac\altaffilmark{3}, K.\ H.\ Kim\altaffilmark{3}, 
        D.\ M.\ Watson\altaffilmark{3}, J. van Cleve\altaffilmark{4},
        J.\ Houck\altaffilmark{5}, K.\ Stapelfeldt\altaffilmark{6},
        M. W. Werner\altaffilmark{6}, G. Rieke\altaffilmark{7}, 
        K.\ Su\altaffilmark{7}, M. Marengo\altaffilmark{8},
	D.\ Backman\altaffilmark{9}, C.\ Beichman\altaffilmark{6}, 
        and G.\ Fazio\altaffilmark{8}
}

\altaffiltext{1}{Spitzer Fellow; NOAO, 950 North Cherry Avenue, 
                 Tucson, AZ 85726; {\sf cchen@noao.edu}}
\altaffiltext{2}{Department of Physics and Astronomy, 
                 University of Missouri, 
                 Columbia, MO 65211; 
                 {\sf lia@missouri.edu}}
\altaffiltext{3}{Department of Physics and Astronomy, 
                 University of Rochester, 
                 Rochester, NY 14627}
\altaffiltext{4}{Ball Aerospace and Technologies Corp., 
                 Boulder, CO 80301}
\altaffiltext{5}{Center for Radiophysics and Space Research,
                 Cornell University, Ithaca, NY 14853-6801}
\altaffiltext{6}{Jet Propulsion Laboratory, Caltech,
                 %California Institute of Technology, 
                 4800 Oak Grove Drive, Pasadena, CA 91109}
\altaffiltext{7}{Steward Observatory, University of Arizona,
                  Tucson, AZ 85721}
\altaffiltext{8}{Harvard-Smithsonian Center for Astrophysics, 
                 60 Garden Street, Cambridge, MA 02138}
\altaffiltext{9}{%SOFIA/SETI Institute,
                 NASA-Ames Research Center, 
                 Moffett Field, CA 94035}

\begin{abstract}
We have obtained \emph{Spitzer} IRS 5.5--35 $\mu$m spectroscopy 
of the debris disk around $\beta$ Pictoris.
In addition to the 10$\mum$ silicate emission feature 
originally observed from the ground, 
we also detect the crystalline silicate emission
bands at 28$\mum$ and 33.5$\mum$. 
This is the first time that the silicate bands 
at wavelengths longer than 10$\mum$ 
have ever been seen in the $\beta$ Pictoris disk.
The observed dust emission is well reproduced
by a dust model consisting of fluffy cometary and 
crystalline olivine aggregates.
We searched for line emission from molecular hydrogen
and atomic [\ion{S}{1}], \ion{Fe}{2}, and \ion{Si}{2} gas 
but detected none. We place a 3$\sigma$ upper 
limit of  $<$\,17$\Mearth$ on the H$_{2}$ S(1) gas mass,
assuming an excitation temperature of $T_{\rm ex}$\,=\,100$\K$.
This suggests that there is less gas in this system than is
required to form the envelope of Jupiter.
We hypothesize that some of the atomic \ion{Na}{1} gas 
observed in Keplerian rotation around $\beta$ Pictoris 
may be produced by photon-stimulated desorption from 
circumstellar dust grains.
\end{abstract}


\section{Introduction}
The A5V star $\beta$ Pictoris (HD 39060 = HR 2020) at a distance 
$d\approx 19.3\pc$ possesses a spectacular edge-on debris disk imaged in 
scattered light and thermal emission that extends to radii $>$1400$\AU$ 
\citep{st84, hol98, gol06}. An age of $\sim$12 Myr has been estimated for 
for the central star based on the properties of late-type members of the
$\beta$ Pic moving group \citep{zsbw01}.
The dust in this system is believed to be replenished 
from a reservoir, such as collisions between parent bodies 
or sublimation of comets because the estimated lifetime 
for the dust under radiation pressure, 
Poynting-Robertson drag, and grain-grain collisions 
is a few orders of magnitude shorter 
than the age of the system \citep{bp93}. 
There may be evidence for the presence
of large bodies in the $\beta$ Pictoris disk 
that perturb dust grains and parent bodies 
and instigate collisional cascades between them. 
A 5$\arcdeg$ warp at radii $<$80 AU has been observed 
in scattered light using STIS on \emph{HST} 
and may be generated by either a brown dwarf 
close to the star (at distances $<$3 AU) or 
a 0.17$\Mjup$ planet at larger distances 
from the star (50$\AU$) \citep{hea00}. 
The $\beta$ Pictoris disk may also possess infalling, 
sublimating, refractory bodies. 
Time-variable, high velocity, non-periodic, 
red-shifted atomic absorption features 
have been observed toward $\beta$ Pictoris 
at ultraviolet (UV) and visual wavelengths 
that vary on timescales as short as hours. 
The velocity of the atoms, typically 100--400$\km\s^{-1}$, 
is close to the free fall velocity at a few stellar radii, 
suggesting that the absorption is produced as stellar photons 
pass through the comae of infalling bodies 
at distances $<$6$\AU$ from the star \citep{vlf98}. 

The origin of the micron-sized dust grains required to 
account for the observed scattered light and thermal emission
is currently not well-constrained. 
The broad 9.7$\mum$ silicate emission feature 
and the narrow 11.3$\mum$ crystalline olivine 
emission feature observed toward $\beta$ Pictoris 
appear grossly similar to those observed toward 
comets Halley, Bradford 1987s, and Levy 1990 XX \citep{kna93}, 
suggesting that the grains may be cometary. 
Models of cometary grains, idealized as fluffy aggregates 
of interstellar silicate cores with carbonaceous 
organic refractory mantles
(and additional ice mantles at distances 
larger than the snow-line), 
in radiative equilibrium with the central star 
are able to reproduce the observed 10$\mum$ silicate feature 
and the spectral energy distribution (SED) 
at IR through millimeter wavelengths \citep{lg98}. 
Spatially-resolved studies of the silicate emission feature, 
obtained using COMICS on the Subaru Telescope, 
suggest that the shape of the 10$\mum$ feature 
changes as a function of position in the disk, 
with large and crystalline grains concentrated 
at the center and small grains located in ring-like bands 
at 6.4, 16, and 29$\AU$ from the central star \citep{oka04}. 
The rings of small grains may be explained by collisions 
between large orbiting bodies. 
High-resolution, mid-IR imaging of $\beta$ Pictoris, 
obtained with TReCs on the Gemini South Telescope, 
has resolved a prominent clump on the south-west 
side of the disk at 52$\AU$ from the central star 
which may also indicate that a recent collision 
has occured in the $\beta$ Pictoris disk \citep{tel05}.

Spatially resolved visual spectroscopy of $\beta$ Pictoris has revealed a disk 
of atomic gas in Keplerian rotation, observed via scattered emission from 
\ion{Fe}{1}, \ion{Na}{1}, \ion{Ca}{2}, \ion{Ni}{1}, \ion{Ni}{2}, \ion{Ti}{1}, 
\ion{Ti}{2}, \ion{Cr}{1}, and \ion{Cr}{2}. The atomic gas possesses a NE/SW 
brightness asymmetry and an inner warp similar to that observed in the dust 
disk although the gas extends to larger heights than the dust \citep{bra04}. 
Estimates of the radiation pressure acting on Fe and Na atoms suggest that 
these species should be not be in Keplerian rotation but should be accelerated 
to terminal outflow velocities $\sim$100s--1000s km/sec \citep{bra04}. 
\cite{lag98} showed that a ring of neutral hydrogen at distance 0.5 AU could 
slow down the radial flow of gas. However, \cite{fbw06} have recently proposed 
that the gas will not be radially accelerated if the carbon is overabundant; 
their model does not require a population of undetected atomic hydrogen. Since 
carbon does not experience strong radiation pressure 
($F_{\rm rad}/F_{\rm grav}$ = $\beta_{\rm carbon}$ $\approx$ 0) 
and also has a large ionization fraction ($\sim$0.5), 
they suggest that Coulomb interactions between \ion{C}{2} and 
other ions reduce the effective radiation pressure on the bulk gas. In this 
case, the ions and neutral atoms in the disk couple together into a fluid, with
an effective radiation pressure coefficient, that is bound to the system and 
that brakes the gas if $\beta_{\rm eff}$ $<$ 0.5. 
In particular, they suggest that 
if the carbon abundance is $>$10$\times$ solar, then all the atomic gas 
will be retained. Measurements of the column density of the stable component of
atomic carbon (with zero velocity relative to the star) via absorption from 
\ion{C}{1} ($^{3}$P) $\lambda$1613 \citep{rob00} and absorption from 
\ion{C}{2}$\lambda$1036 and \ion{C}{2}$^{*}$ $\lambda$1037 superimposed on
chromospheric \ion{O}{6} $\lambda$1038, suggest that the bulk of the atomic gas
is composed of carbon with a C/Fe abundance ratio that is 16 times the solar 
value and an ionization fraction of 0.5 \citep{rob06}. 

We report the results of a \emph{Spitzer} IRS study of the dust and gas around 
$\beta$ Pictoris, building on the model for the composition and spatial 
distribution of the grains by \cite{lg98} and placing 3$\sigma$ upper limits on
the line emission from H$_{2}$ S(2), S(1), S(0) and [\ion{S}{1}], \ion{Fe}{2}, 
and \ion{Si}{2} gas. Based upon the similarity in the spatial distribution of 
the dust and gas observed in scattered light, we hypothesize that the dust and 
gas in this system are physically associated and that the observed gas is 
secondary; it has been produced from circumstellar material since the star 
formed. We quantitatively show that the observed \ion{Na}{1}, seen in Keplerian
rotation around the star, can be produced via photon-stimulated desorption in a
time that is shorter than the residence time of the gas in the disk.

\section{Observations}
We obtained IR Spectrograph (IRS) \citep{hou04} spectral 
mapping mode observations of $\beta$ Pictoris 
using the Short-Low 
(SL; 5.2--14 $\mum$; $\lambda/\Delta\lambda \sim 90$), 
Short-High 
(SH; 9.9--19.6 $\mum$; $\lambda/\Delta\lambda \sim 600$), 
and Long-High 
(LH; 18.7--37.2 $\mum$; $\lambda/\Delta\lambda \sim 600$)
modules on the \emph{Spitzer Space Telescope} \citep{wer04}. 
The SL2 slit has a size 3.6$\arcsec\times$57$\arcsec$; 
the SL1 slit has a size 3.7$\arcsec\times$57$\arcsec$; 
the SH slit has a size 4.7$\arcsec\times$11.3$\arcsec$; 
the LH slit has a size 11.1$\arcsec\times$22.3$\arcsec$. 
Both the low and high resolution observations 
were made with the spectrograph long slits aligned 
along the position angle of $\beta$ Pictoris disk to within 5$\arcdeg$. 

The SL2 (AOR key: 8972288) and SL1 (AOR key: 8972544) spectra were obtained on 
16 November 2004 by stepping the slit in 7 and 11 1.8$\arcsec$ intervals 
perpendicular to the disk, respectively, with the center position centered on 
the disk. Each SL2 and SL1 map position was observed using 9 and 14 cycles, 
respectively, of 6 second ramps. The SH (AOR key: 4879616) and LH (AOR key: 
4876800) spectra were obtained on 15 December 2003 and 4 March 2004, 
respectively, by stepping the slit in 3 2.4$\arcsec$ and 4.8$\arcsec$ intervals
perpendicular to the disk, respectively, with the center position centered on 
the disk. Each SH and LH map position was observed using 25 and 24 cycles, 
respectively, of 30 and 14 second ramps. High accuracy peak-up, with the blue 
(13.3--18.7$\mum$) array, on the nearby star HD 38891 
(located 12$\arcmin$ north of $\beta$ Pictoris), 
was used to center the disk in the SL2, SL1, and SH slits;
high accuracy peak-up, with the red array (18.5--26$\mum$), 
on HD 38891 was used to center the disk in the LH slit. 

The raw data were processed into calibrated (flat-fielded, stray light 
corrected) 2-dimensional spectra with version S11.0.2 of the SSC pipeline. We 
removed masked and rogue pixels from the basic calibrated data by interpolating
between their nearest neighbors, coadded the resulting images, and extracted 
1-dimensional spectra using the IRS team's SMART program \citep{hig04}. Since 
separate nod observations of the sky were not made when $\beta$ Pictoris was 
observed, we approximated the SL sky background as a constant at each 
wavelength and subtracted this value from each pixel in the spatial direction
across the source. Since the SL slits are long enough to sample the
PSF in the spatial direction, we fit 1-dimensional Gaussians to the SL spectra 
in the spatial direction to determine the positions of the source in the slits.
We extracted our low resolution spectra by summing over a window whose width 
varied as a function of wavelength to account for the changing size of the 
point spread function and was centered on the source position. Finally, we 
multiplied the resulting spectra with an $\alpha$ Lac relative spectral 
response function (RSRF), obtained by dividing an $\alpha$ Lac template 
spectrum with an observed $\alpha$ Lac SL spectrum \citep{jur04}. Since point
sources fill the high resolution slits, we extracted our high resolution 
spectra (without background subtraction) by summing over all of the pixels in 
the SH and LH slits. The extracted high resolution spectra were multiplied by a
$\xi$ Dra high resolution RSRF. 

We constructed a spectrum that is consistent with the LH slit, the largest slit
we used to observe $\beta$ Pictoris, centered on the disk. Since the source is
extended in our IRS observations, we added multiple SL measurements together 
to construct an aperture equal to the LH aperture. Since our SH map was not 
large enough to cover the LH central slit position, we scaled our SH spectrum 
by a factor of 1.068 to match the SL and LH spectra, using the LL spectrum at 
5.1939 $\mu$m to 9.9232 $\mu$m, the SH spectrum at 9.9461 $\mu$m to 19.3179 
$\mu$m, and the LH spectrum at 19.2636 $\mu$m to 35.9892 $\mu$m. The final 
spectrum of $\beta$ Pic (shown in Figure \ref{fig:sedobs}) is that of a 
11$\arcsec$$\times$22$\arcsec$ region centered on the star and oriented along 
the disk. 

The signal:noise in our spectrum is very high. Over spans of many spectral 
resolution elements, the sensitivity of our spectrum is limited by 
flat-fielding errors to 1-2\% of the continuum flux density. Over smaller 
spans, $\le$ 10 spectral resolution elements or so, the sensitivity is limited 
only by photon noise from the background and target, and is significantly 
better: 0.1-0.5\% of the continuum flux density in SH and LH, and 0.05-0.2\% in
SL, consistent with the point-to-point scatter in the spectrum over such small 
spans, and (perforce) consistent with the {\it Spitzer}-IRS facility 
sensitivity tool, SPEC-PET. Thus, every spectral feature visible in our $\beta$
Pic spectrum that is at least as wide as the spectral resolution is real, and 
the upper limits on weaker spectral lines are significantly smaller than 1\% of
the continuum. For simplicity we have calculated such upper limits directly 
from the point-to-point variation of the continuum-subtracted spectrum.  

\section{Dust Properties}
\begin{figure}
\figurenum{1}
\plotone{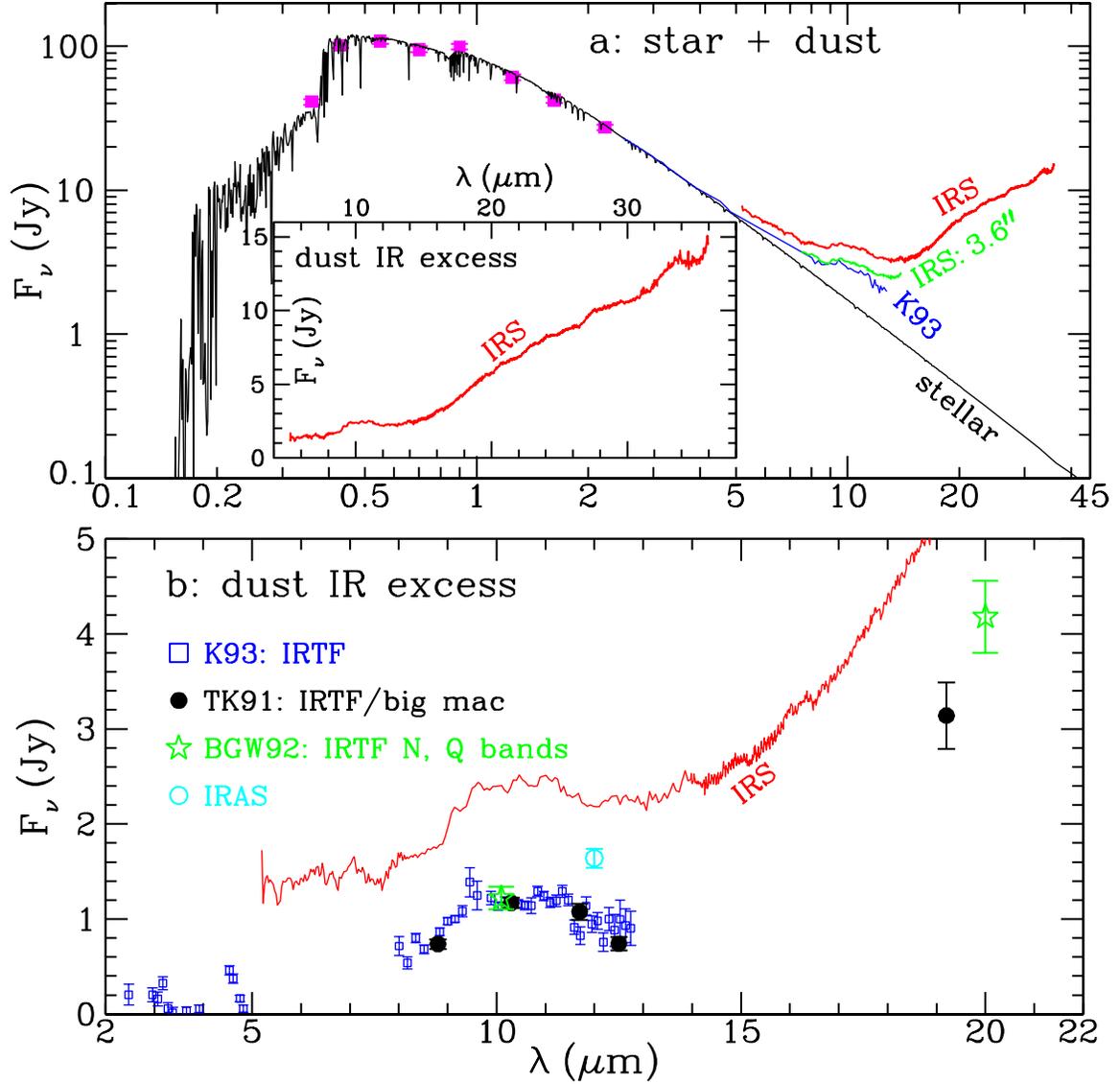}
\caption{
\label{fig:sedobs}
(a) \emph{Spitzer} IRS SL, SH, and LH spectrum 
of $\beta$ Pictoris (red line) 
with the PHOENIX atmospheric spectrum (black line) overlaid.
The inserted panel illustrates the 5--36$\mum$
stellar-subtracted \emph{Spitzer} IRS dust excess emission.
Also plotted are the UBVRIJHK stellar photometry 
(magenta squares), the ground-based spectrometry
of \cite{kna93} (labeled ``K93''), and the IRS spectrum 
in the inner 3.6$^{\prime\prime}$ 
[corresponding to $\sim$\,35$\AU$; approximately the same
beam size as that of \cite{kna93}] disk (green line).
(b) Comparison of the stellar-subtracted \emph{Spitzer} 
IRS dust excess emission (red line) with 
the IRAS photometry (cyan open circles),
the {\it Big Mac}/IRTF photometry (black filled circles),
the IRTF $N$ band and $Q$ band photometry (green open stars),
and the IRTF 2.6--13.5$\mum$ spectrometry 
of \cite{kna93} (blue open squares). Numerous hydrogen 
absorption features can be seen between 5 and 13 $\mu$m; 
their depths are consistent with that expected
for a star with A5V photosphere at spectral resolution of 
the IRS.
}
\end{figure}

Figure \ref{fig:sedobs} plots the 5.2--36$\mum$ IRS spectrum
and the PHOENIX model stellar atmospheric spectrum 
with effective temperature $T_\star\,=\,8000\K$
and surface gravity log\,$g$ = 4.2. 
Figure \ref{fig:sedobs} clearly shows that 
the dust excess emission above the stellar
atmospheric radiation is even seen at wavelengths
as short as $\lambda\sim$\,5--8$\mum$,
indicating that there may exist a population of 
hot dust ($\sim$\,400--600$\K$)
at a distance of $<$10$\AU$ from the central star
[see Figure 1 of \cite{lg98}].

Mostly prominent in Figure \ref{fig:sedobs} 
are the broad 10$\mum$ feature (which reveals 
the presence of both amorphous silicates and 
crystalline silicates in the $\beta$ Pictoris disk), 
and the 28$\mum$ and 33.5$\mum$
crystalline olivine emission features
(see the inset in Figure \ref{fig:sedobs}a)
which closely resemble the \cite{lg98} model 
spectrum (see their Figure 6c).\footnote{%
  Crystalline pyroxene Mg$_{x}$Fe$_{(1-x)}$SiO$_3$
  grains also have a rich set
  of features in the mid-IR. But their peak wavelengths
  and relative feature strengths are inconsistent 
  with the IRS spectrum (e.g. see Chihara et al.\ 2001).    
  Olivine Mg$_{2x}$Fe$_{2(1-x)}$SiO$_4$ is a mixture 
  of forsterite Mg$_2$SiO$_4$ and fayalite Fe$_2$SiO$_4$ 
  with a mixing proportion $x$. 
  In this work as well as in Li \& Greenberg (1998) we
  adopt the refractive indices of Mg$_{1.8}$Fe$_{0.2}$SiO$_4$ 
  (Mukai \& Koike 1990). Although in principle
  it is possible to infer the Mg:Fe ratio from the locations 
  and strengths of the IR features 
  (e.g. see Koike et al.\ 1993, 2003, Fabian et al.\ 2001), 
  neither the IRS spectrum reported here reveals a large number
  of crystalline silicate features nor does it cover a sufficiently
  broad wavelength range for constraining the exact crystal composition.   
  As shown in Figure 2, the dust model consisting of
  crystalline Mg$_{1.8}$Fe$_{0.2}$SiO$_4$ olivine reproduces
  the IR features reasonably well. 
  }
This is the first time that the silicate bands 
at wavelengths longer than 10$\mum$ have ever
been detected in the $\beta$ Pictoris disk.

\begin{figure}
\figurenum{2}
\plotone{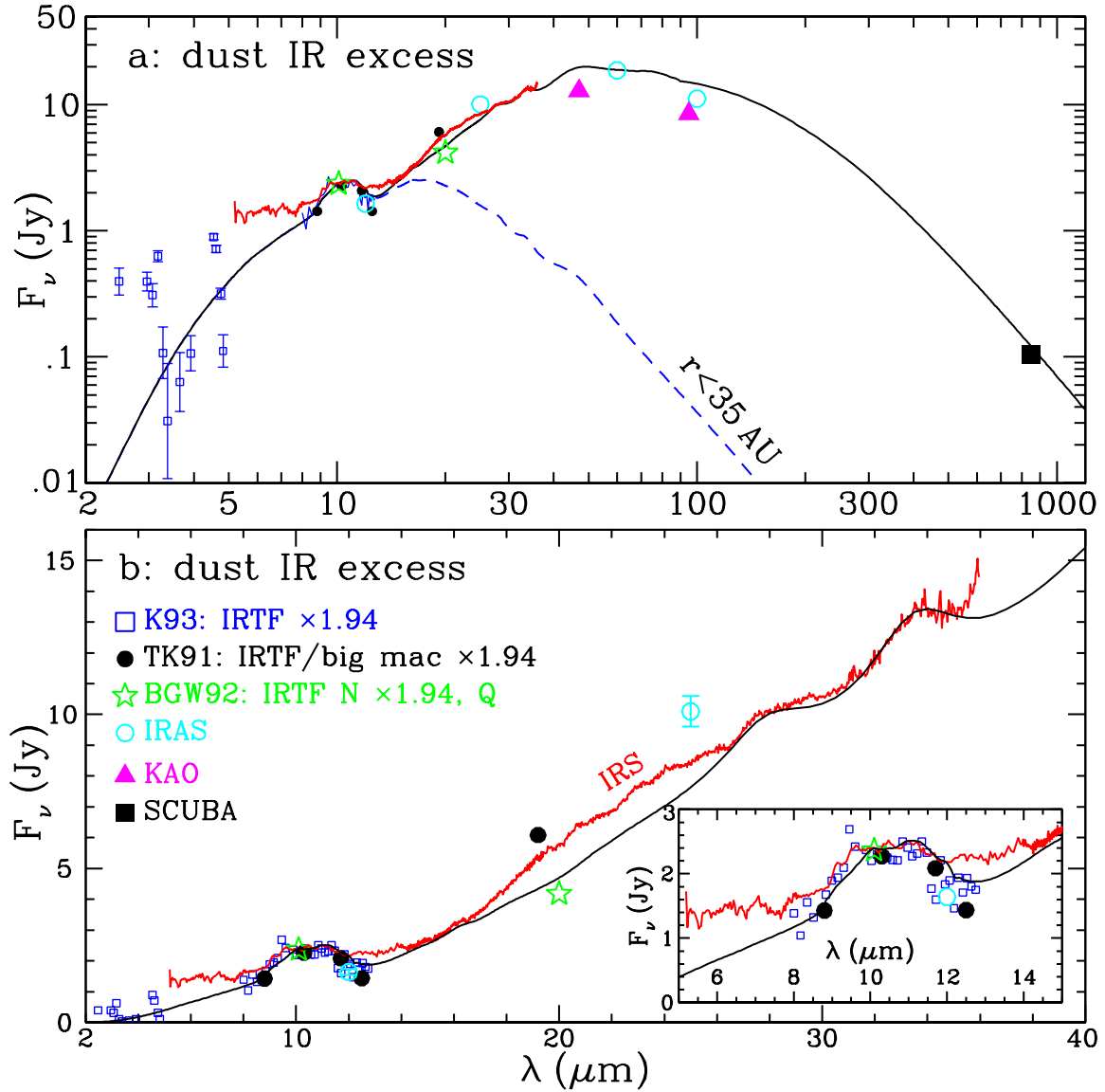}
\caption{
\label{fig:sedmod}
Comparison of the model (black line) to 
(a) the observed dust IR emission spectral energy distribution 
and (b) the \emph{Spitzer} IRS spectrum (red line).
Also plotted is the model-predicted emission 
in the inner 35$\AU$ disk.
Blue open squares: the IRTF 2.6--13.5$\mum$ spectrometry 
of \cite{kna93} increased by a factor of 1.94.
Black filled circles: the {\it Big Mac}/IRTF photometry
of \cite{tk91} increased by a factor of 1.94.
Green open stars: the IRTF $N$ band and $Q$ band photometry
of \cite{bgw92} with the $N$ band increased by a factor of 1.94.
Cyan open circles: the IRAS photometry of \cite{gil86}.
Magenta filled triangles: the KAO photometry of \cite{har96}.
Black filled squares: the 850$\mum$ SCUBA photometry of \cite{hol98}.
With this increase (of a factor of 1.94), all observational data
agree with each other very well. 
The inserted panel in (b) illustrates the model fit
to the 10$\mum$ silicate features. 
}
\end{figure}

For comparison, we also show in Figure \ref{fig:sedobs} 
the IRAS broadband photometry of \cite{gil86},
the photometry of \cite{tk91}
at 8.8, 10.3, 11.7 and 12.5$\mum$
obtained using the {\it Big Mac} spectral filters
at the NASA {\it Infrared Telescope Facility} (IRTF), 
the IRTF $N$ band (10.1$\mum$) and $Q$ band (20$\mum$) 
photometry of \cite{bgw92},
the IRTF 2.6--13.5$\mum$ intermediate resolution
($\lambda/\Delta\lambda\approx 50$) spectrometry 
of \cite{kna93}, 
and the {\it Kuiper Airborne Observatory} (KAO)
47 and 95$\mum$ photometry of \cite{har96}.
%Since these observations were made 
%with different aperture sizes, 
While the photometry of \cite{tk91}
and the IRTF $N$ band photometry of \cite{bgw92}
agree with the spectrometry of \cite{kna93} very well,
they are below the IRS spectrum by a factor of $\sim$\,1.94
(see Figure \ref{fig:sedobs}b and  Figure \ref{fig:sedmod}b).  
This difference is
much larger than the calibration uncertainty of IRS and
is attributed to the fact that the Spitzer apertures 
($3.6^{\prime\prime}\times 57^{\prime\prime}$ for SL,
$4.7^{\prime\prime}\times 11.3^{\prime\prime}$ for SH,
$11.1^{\prime\prime}\times 22.3^{\prime\prime}$ for LH)
are much larger than the apertures of ground-based observations 
[e.g. $3.7^{\prime\prime}$ diameter of \cite{kna93},  
$3.9^{\prime\prime}\times 4.2^{\prime\prime}$ of \cite{tk91}, 
$7.8^{\prime\prime}$ diameter of \cite{bgw92}],
and therefore Spitzer detects more flux from extended
material than in previous observations.\footnote{%  
   Indeed, as shown in Figure 1a, the SL point-source-extracted 
   spectrum centered on the star (with a beam diameter of
   $\sim$\,3.6$^{\prime\prime}$, approximately the same 
   as \cite{kna93}'s spectroscopic observation) is close
   to that of \cite{kna93}.
%   {\bf Dana: could you write a few words to answer
%   the referee's question ...} 
   }

The nature (e.g. size, composition, and morphology) 
of the dust and its spatial distribution in protoplanetary 
and debris disks is mostly revealed through the interaction 
of the dust with the electromagnetic radiation of 
the central star: absorbing, scattering and polarizing 
the stellar radiation and re-radiating 
the absorbed UV/visible photons at longer wavelengths,
ranging from the IR to submillimeter and millimeter.
Since the disk around $\beta$ Pictoris subtends 
more than 100$\arcsec$, it has been studied in great detail.
There exists a vast variety of observational data 
for the dust in this disk (see \cite{mann06}),
including (1) imaging observations of scattered light
\citep{st84, pb87, gdc93, lde93, 
kj95, mou97, hea00, kal00, gol06};
(2) imaging observations of dust thermal emission 
in the mid-IR 
\citep{lp94, pla97, hei99, wah03, tel05} 
and submillimeter \citep{hol98}; 
(3) mid-R narrow band photometry \citep{tk91},
as well as broadband photometry 
in the mid-IR \citep{gil86, bgw92}
and millimeter \citep{chi91, lis03}; 
(4) optical \citep{gsw91, wsg95} and near-IR \citep{tam06} 
imaging of polarized light; and 
(5) mid-IR spectroscopy \citep{ait93, kna93, wbz03, oka04}.

To accurately describe the dust grains 
in the $\beta$ Pictoris disk, 
all of the aforementioned observational 
data must be modeled {\it simultaneously}.
%So far, there lacks such a comprehensive modeling,
%although extensive modeling efforts have been 
%performed in literature. All existing work focus 
%on particular observations of the disk -- 
%either scattered light \citep{abp89, kj95, aug01, gol06}, 
%polarization \citep{vk99, kkm00, tam06}, 
%or dust thermal emission \citep{chi91, bgw92, ait93, 
%kna93, lg98, hei99, sdw04}.\footnote{%
%  \cite{pla97} modeled both the optical scattered light 
%  and the mid-IR emission; however, they simply adopted 
%  the Henyey-Greenstein phase function instead of 
%  calculating the scattering parameters of the dust.
%  } 
%
While recognizing the importance of a comprehensive,
simultaneous modeling of all of the $\beta$ Pictoris 
observations, in this paper we present a simple model 
focusing on our \emph{Spitzer} IRS spectrum.
We defer the comprehensive modeling of the dust 
around $\beta$ Pictoris that attempts to simultaneously 
fit all of the observed data
to a separate paper (A. Li et al.\ 2007, in preparation).

We model the \emph{Spitzer} IRS spectrum of $\beta$ Pictoris 
building on the work of \cite{lg98} which includes 
(1) a population of cometary dust, composed of low-density 
porous aggregates of amorphous silicate core-carbonaceous 
mantle grains,\footnote{% 
  The mass ratio of the silicate core to the carbon mantle
  is assumed to be $m_{\rm carb}/m_{\rm sil} \approx 0.7$
  based on the cosmic abundance consideration
  (see Appendix A of Li \& Lunine 2003).
  } 
and (2) a population of crystalline olivine aggregates
with the same size and spatial distributions 
as the amorphous cometary grains
(but actually we do not require crystalline dust 
outside $\sim$\,60$\AU$ from the star 
since at distances $>$\,60$\AU$ from the star
silicate grains are too cold to emit at
the characteristic mid-IR bands). 

For the amorphous cometary and crystalline olivine 
fluffy aggregates, we assume a vacuum volume fraction 
$P$\,=\,0.90 (see \S2 of Li, Lunine, \& Bendo 2003 for justification) 
and approximate the porous grains as spheres with radii $a$. 
We assume a power-law dust size distribution 
$dn(a)/da \propto a^{-\alpha}$, 
with a minimum grain radius $\amin=1\mum$, 
and a maximum grain radius $\amax=10000\mum$
(see Li, Lunine, \& Bendo 2003 for justification). 

Similar to \cite{ac97}, we assume a modified power-law 
spatial distribution for the dust, 
$n(r) \propto 
1/\left[f\left(r/\rp\right)^{-1}+\left(r/\rp\right)^{\gamma}\right]$.
This functional form, peaking at 
$\rp\times\left(f/\gamma\right)^{1/\left(\gamma+1\right)}$,
behaves like a power-law $n(r) \propto r^{-\gamma}$ 
at large distances ($r>\rp$).
At $r<\rp$, the $\left(r/\rp\right)^{-1}$ term
dominates, approximating an increasing profile 
in the inner, dust-relatively-depleted region. 
We take $\rp=100\AU$ and $\gamma=2.7$ as derived 
from scattered light modeling \citep{abp89}.
The inner boundary $\rmin$ of the disk is taken to 
be the location inside which silicate dust sublimates.  
For the $\beta$ Pictoris disk, micron-sized silicate 
grains possess an equilibrium temperature
$T\approx 1500\K$ at $r\approx 0.2\AU$; 
therefore, we take $\rmin=0.2\AU$. 
We take $\rmax$\,=\,2000$\AU$. 

In modeling the dust IR emission,
we -- with all other parameters 
(i.e. $\rp$, $\gamma$, $\rmin$, $\rmax$) pre-chosen --
are therefore left with only 2 free parameters:
$\alpha$ -- the dust size distribution power-index,
and $f$ which determines the amount of dust 
in the inner disk region $r<\rp$
[we note that \cite{ac97} took $f$\,=\,1]. 
It is found that, with $\alpha\approx 3.2$
and $f \approx 0.15$, we are able to obtain 
a reasonably good fit to
the overall SED and the Spitzer IRS spectrum 
including the silicate emission features 
at 10, 28 and 33.5$\mum$
(see Figure \ref{fig:sedmod}).\footnote{%
  The amount of dust in the inner disk ($r<\rp$)
  is inverse proportional to $f$.
  With $f$\,=\,0.15 and $\gamma$\,=\,2.7,
  the dust spatial distribution peaks at 
  $0.46\,\rp\approx 46\AU$. At a first glance,
  this appears inconsistent with the scattered light images
  which suggest a dust spatial distribution slope
  change at $\sim$\,80--120\,AU 
  (e.g. see \cite{gdc93,gol06}).
  However, the dust spatial distribution from
  46\,AU ($\approx\frac{1}{2}\rp$) to $\rp$ is rather
  flat: $n(r)$ only changes by a factor of $\approx$2.3
  from $\frac{1}{2}\rp$ to $\rp$, while it drops much more 
  steeply from $\rp$ to 2\,$\rp$ (by a factor of 
  $\approx$5.8). In a subsequent paper (A. Li et al.\ 2007,
  in preparation), we will investigate in detail whether
  the present dust spatial distribution
  is able to reproduce the scattered light images.
  }
%except the deficiencies at $\lambda\sim$\,5--8$\mum$ 
%and $\lambda\sim$\,17--24$\mum$ 
%[see \S6 and A. Li et al.\ (2007, in preparation)].
The required dust masses are approximately
$1.63\times 10^{27}\g\approx 0.27\,{\rm M}_\earth$ 
and $4.89\times 10^{25}\g\approx 0.0082\,{\rm M}_\earth$ 
for the amorphous and crystalline components, respectively.

As shown in Figure \ref{fig:sedmod}b,
the model is somewhat deficient 
at $\lambda\sim$\,5--8$\mum$ 
and $\lambda\sim$\,17--24$\mum$. 
This is due to the oversimplified dust spatial 
distribution function adopted above.
By including 2 rings or clumps of dust 
at $r<10\AU$ and at $r\sim$\,20--30$\AU$
as implied by \cite{tel05} and \cite{oka04},
it is expected that the model will account for 
the excess emission at 5--8$\mum$ and 17--24$\mum$
(see A. Li et al.\ 2007, in preparation).
The oversimplification of the assumed dust spatial 
distribution is also reflected by the model-predicted
IR emission from the dust within a 3.6$^{\prime\prime}$
diameter (corresponding to $\sim$\,35$\AU$; see Figure 2a):
the model-predicted emission from the inner 35$\AU$ region
accounts for almost all the observed $\lambda$\,$<$\,15$\mum$ 
emission for the entire disk. Although this is consistent with
the previous spatially resolved mid-IR spectroscopy 
which show that 10$\mum$ silicate emission 
originates from the inner $\sim$\,20$\AU$ 
\citep{wbz03, oka04}, it appears to contradict the IRS
detection of an appreciable amount of silicate emission
from the outer $\sim$\,35$\AU$ (see Figure 1a).
This will be investigated in detail in a subsequent 
paper (see A. Li et al.\ 2007, in preparation).

\section{Gas Mass Upper Limits}
Core-accretion models suggest that giant planets accrete their gaseous 
envelopes on timescales between 1 and 20\,Myr. Therefore the $\beta$ Pictoris 
disk, with an age of $\simali$12\,Myr, is an excellent source to search for 
bulk gas. If the gas has a solar composition, then the bulk gas is expected to 
be hydrogen. \emph{ISO} observations indicated line emission from H$_{2}$ S(0) 
at 28.2$\mum$ and S(1) at 17.0$\mum$, suggesting that the disk possesses 
54$\Mearth$ warm H$_{2}$ with an excitation temperature 
$T_{\rm ex}$\,=\,110$\K$ \citep{thi01}. However, apparently conflicting 
\emph{FUSE} observations constrain the circumstellar H$_{2}$ Lyman series 
absorption and place a 3$\sigma$ upper limit of 
$N({\rm H}_2) \leq 10^{18}\cm^{-2}$ on the H$_{2}$ column density, 
significantly lower than (5--500)$\times 10^{20}\cm^{-2}$ expected if the 
\emph{ISO}-detected H$_{2}$ were uniformly distributed in an edge-on disk in 
the beam \citep{lde01}, an assumed geometry that is consistent with 
observations of the atomic gas. Detailed studies of gas drag on the dust 
dynamics suggest that the gas:dust ratio is less than 1 (or $<$0.4 $M_{\earth}$
molecular hydrogen exists in the disk). For example, if the disk possessed 
40 M$_{\earth}$ gas, then small grains would collect at distances $>$200 AU and
would increase the scattered light surface brightness by more than a factor of
10 \citep{ta05}. Searches for \ion{H}{1} 21 cm emission constrain the mass of 
atomic hydrogen $M_{\rm HI} \leq 0.5\Mearth$ \citep{fre95}; recent chemical 
models constrain the total mass of hydrogen $\leq$15 M$_{\earth}$ including the
molecular component based on these observations \citep{kfc07}.

\begin{figure}
\figurenum{3}
\plottwo{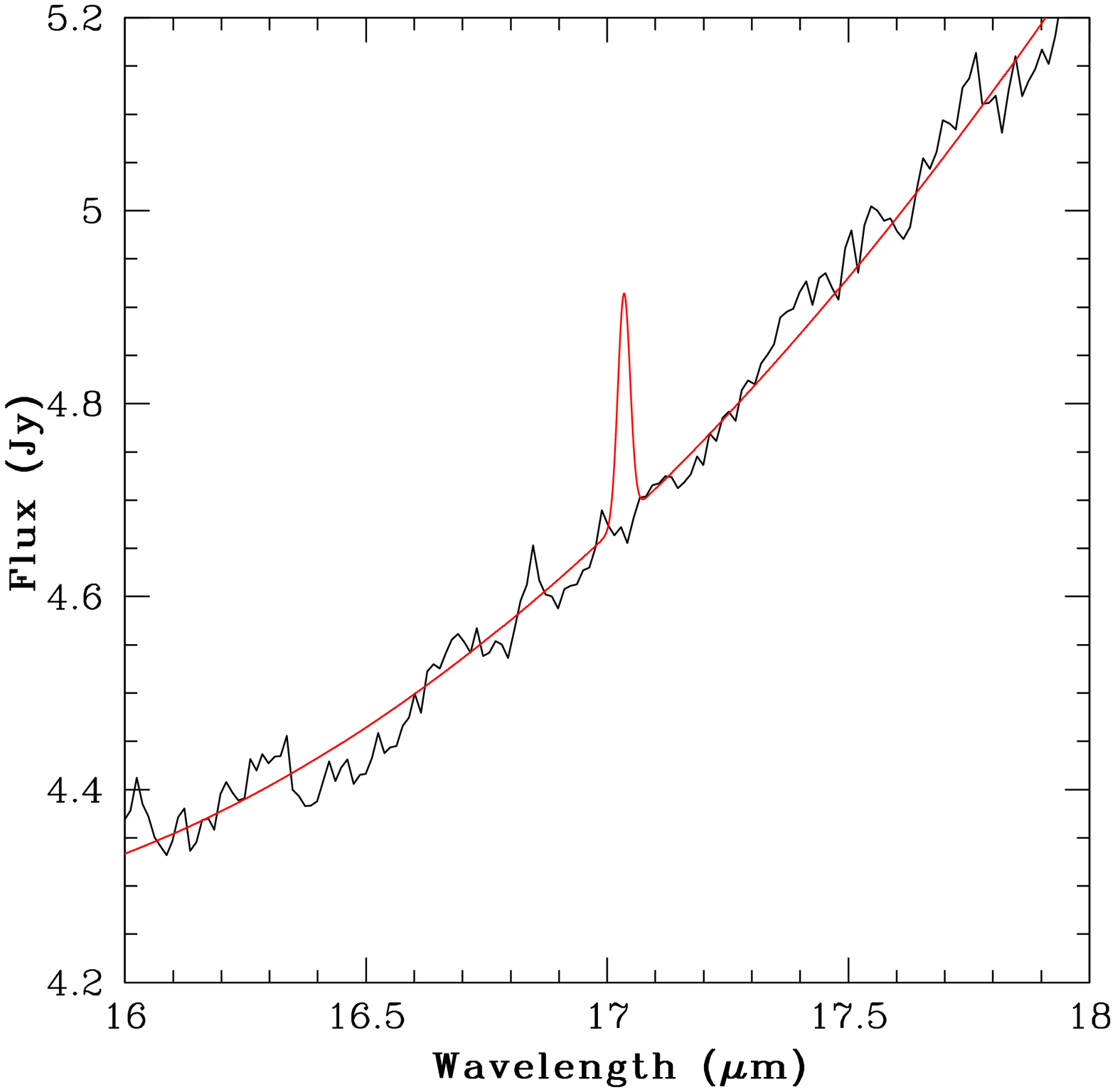}{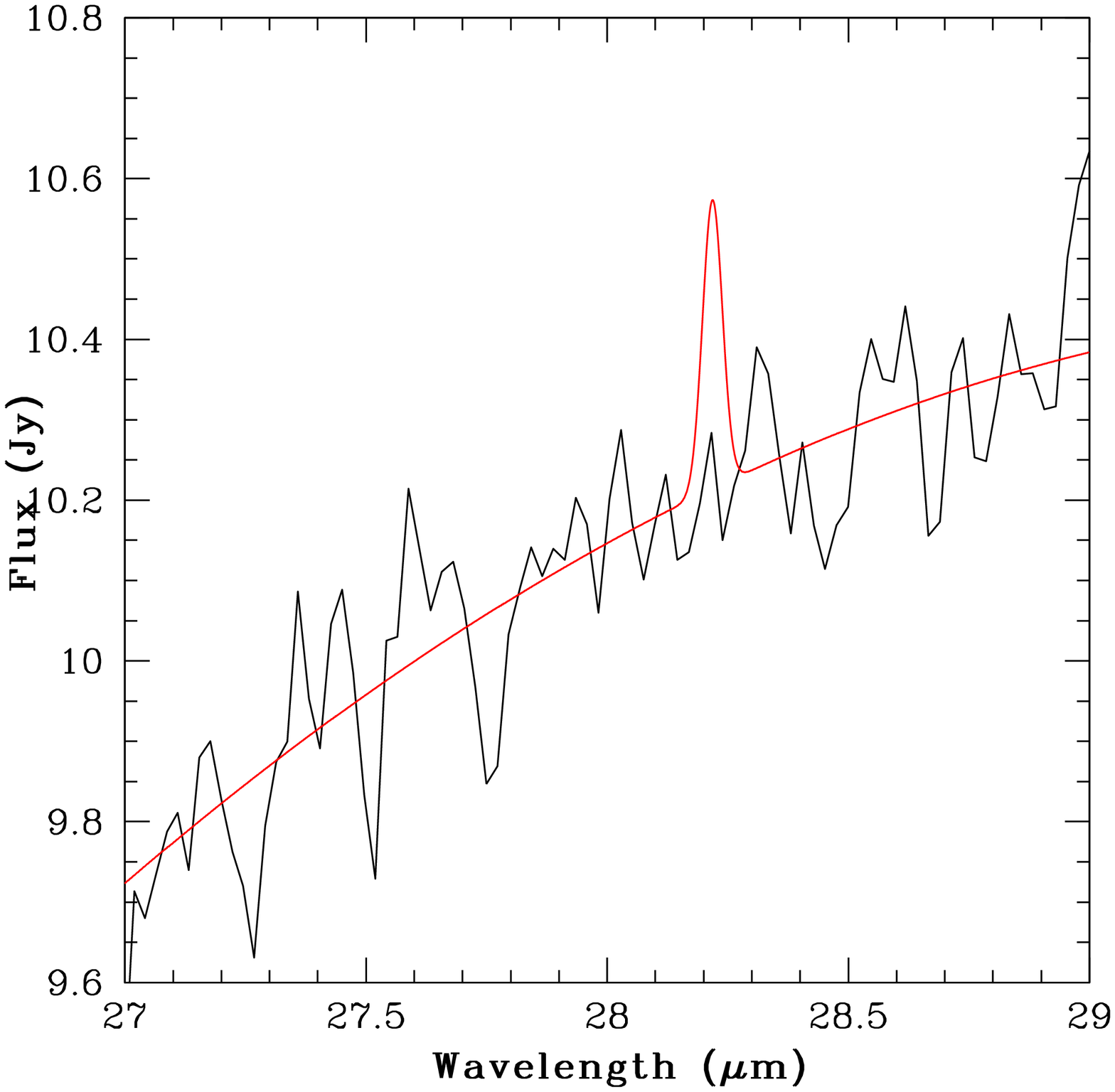}
\caption{
\label{fig:H2_spec}
(a) Portion of the observed \emph{Spitzer} SH spectrum 
of $\beta$ Pictoris around H$_{2}$ S(1) 
with a model for unresolved H$_{2}$ emission 
(overplotted in red) inferred from \emph{ISO} detection. 
(b) Same as (a) but for the \emph{Spitzer} LH spectrum 
around H$_{2}$ S(0).}
\end{figure}

We searched for emission from H$_{2}$ S(2), S(1), S(0) and [\ion{S}{1}], 
\ion{Fe}{2}, and \ion{Si}{2} but did not detect any of these species. We place 
3$\sigma$ upper limits on their line fluxes toward $\beta$ Pictoris (see Table 
1). Our 3$\sigma$ upper limit on the H$_{2}$ S(1) line flux is 
$<$1.2$\times 10^{-14}\erg\s^{-1}\cm^{-2}$, a factor of $\simali$6.4 times 
lower than the reported \emph{ISO} detection 
(see Figure \ref{fig:H2_spec}). 
Although the \emph{Spitzer} IRS SH slit 
is a factor of two shorter than the \emph{ISO} SWS 
slit (14$\arcsec \times$27$\arcsec$), our upper limits effectively constrain 
the H$_{2}$ line emission because the warm, bulk H$_{2}$ is expected to be 
located at radii $<$100$\AU$ ($<$6$\arcsec$). %Detailed chemical models of
%intermediate-age disks by \cite{gh04} suggest that [\ion{S}{1}] may set the
%most stringent gas mass upper limit of the \emph{Spitzer} infrared line 
%diagnostics. Our 3$\sigma$ upper limit on the [\ion{S}{1}] line flux is 
%$<$8.6$\times 10^{-14}\erg\s^{-1}\cm^{-2}$.

Converting 3$\sigma$ upper limits on the line fluxes from any species into gas 
masses depends sensitively on the assumed gas temperature. Detailed models of 
the heating and cooling of molecular and atomic gas via gas-grain collisions, 
cosmic rays, line emission, etc. have been used to infer the temperature and 
chemical structure of the $\beta$ Pictoris disk. However, these models depend 
on the initial gas:dust ratio or gas mass assumed. For example, a 
2$\Mearth$ disk (with an interstellar gas:dust ratio of $\sim$100) heated 
primarily by gas-grain collisions may possess gas as warm as 100--150$\K$ at 
intermediate heights at distances of 300--500$\AU$ where [\ion{O}{1}] fine 
structure emission at 63.2$\mum$ is the dominant coolant; while, a 0.2$\Mearth$
disk may possess gas as warm as 300$\K$ at lower heights and similar radii in 
the disk where [\ion{O}{1}] and H$_{2}$ rotational/vibrational emission are the
dominant coolants \citep{kvz01}. Estimates of the total gass mass, inferred
from the measured column densities of atomic species and the scattered light 
gas density profile of \ion{Na}{1}, suggest that the $\beta$ Pictoris disk 
contains $\sim$7.4$\times$10$^{-4}$ $M_{\earth}$ measured gas, corresponding to
a gas:dust ratio $\sim$0.019, significantly less than assumed in many detailed
chemical models. Even if the gas has a solar hydrogen abundance relative to the
heavy elements, then the disk possesses $\sim$3.7$\times$10$^{-3}$ $M_{\earth}$
gas, corresponding to a gas:dust ratio $\sim$0.093 (A. Roberge, private 
communication), still significantly less than the gas:dust ratios of 100 or 10
assumed in these models. 

\begin{figure}
\figurenum{4}
\plotone{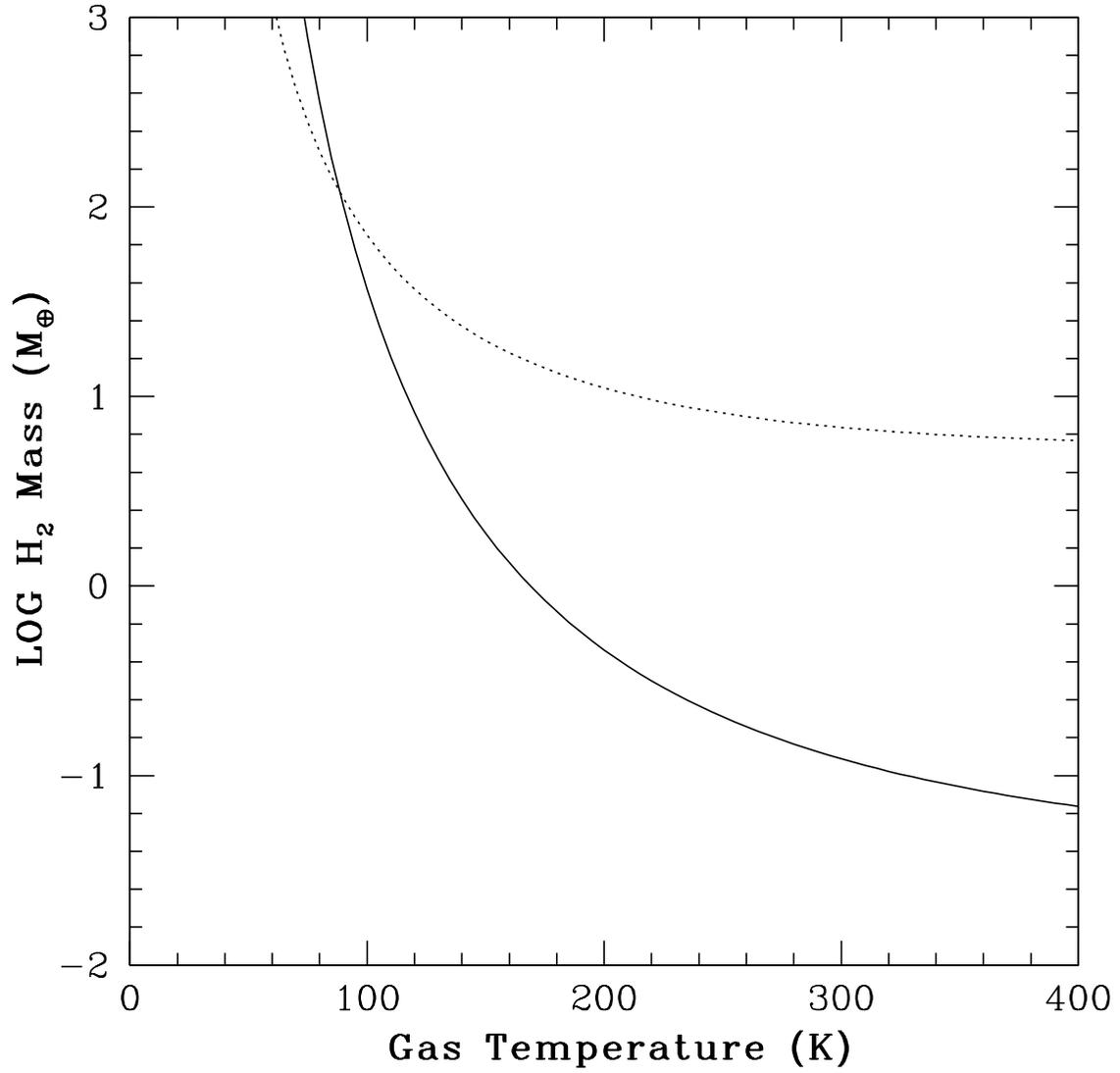}
\caption{
\label{fig:H2_mass}
Three $\sigma$ upper limits on the hydrogen gas mass 
as a function of disk gas temperature, 
estimated from the line fluxes of
%[\ion{S}{1}] assuming an interstellar abundance (dashed line),
H$_{2}$ S(1) (solid line), and H$_{2}$ S(0) (dotted line).} 
\end{figure}

To estimate an upper limit on the mass of H$_{2}$ based on our S(1) line flux 
upper limit, we must assume a gas excitation temperature. Gas temperatures can 
be inferred from (1) detailed models of gas in thermal balance that calculate 
the composition and density structure of disks in addition to the temperature 
structure (as described above) or (2) observations if multiple electronic 
transitions are observed. Far-UV observations of $\beta$ Pictoris have 
constrained the bulk circumstellar gas temperature. Analysis of the \ion{C}{1} 
($^{3}$P) $\lambda$1613 and $\lambda$1561 multiplets suggest that the 
excitation temperature of the stable \ion{C}{1} component, $T_{\rm ex}$\,=\,
50--100$\K$ \citep{rob00}. Since bulk hydrogen with a gas:dust ratio of 100 has
not been directly detected (as assumed for the Kamp \& van Zadelhoff models), 
we rely on the measured \ion{C}{1} excitation temperature to act as a guide for
the bulk gas kinetic temperature $T_{\rm kin}$\,=\,50--100$\K$. Therefore, we 
place 3$\sigma$ upper limits on the mass in each of the listed species assuming
gas temperatures $T_{\rm ex}$\,=\,50 and 100$\K$ (see Table 1). The total flux 
produced by $N$ atoms or molecules
\begin{equation}
F = \frac{h\nu N \chi_{u} A_{ul}}{4 \pi d^{2}}
\end{equation}
where $d$ is the distance to the star, $E = h \nu$  is the energy of the 
radiated photons, $\chi_{u}$ is the fraction of atoms or molecules in level 
$u$, and $A_{ul}$ is the transition probability. Since the temperature of any 
H$_{2}$ is uncertain, we plot the 3$\sigma$ upper limits on the H$_{2}$ mass 
obtained from our S(1) and S(0) line fluxes %and from our \ion{S}{1} line flux 
%(assuming that the gas has an interstellar composition) 
as a function of temperature (see Figure \ref{fig:H2_mass}). 
If $T_{\rm gas}$\,=\,100\,K, we estimate that $<$17$\Mearth$ H$_{2}$ 
%or $<$0.17$\Mearth$ 
hydrogen remains in the disk from our H$_{2}$ S(1) %and [\ion{S}{1}] 
upper limit.

The nondetection of H$_{2}$, [\ion{S}{1}], \ion{Fe}{2}, and \ion{Si}{2} 
emission toward $\beta$ Pictoris is consistent with \emph{Spitzer} observations
of other 10--30\,Myr old disks. \emph{ISO} observations of the 20\,Myr old 
dusty A1V star 49 Ceti have indicated line emission from H$_{2}$ S(0) at 
28.2$\mum$, suggesting that the 49 Cet disk possesses $>$110$\Mearth$ warm 
H$_{2}$ with an excitation temperature $T_{\rm ex} < 100\K$ \citep{thi01}. 
However, high resolution \emph{Spitzer} IRS observations place 3$\sigma$ upper 
limits on the H$_{2}$ S(0) line flux that are a factor $\sim$9 lower than 
the reported detection \citep{che06}. A recent \emph{Spitzer} IRS search for 
H$_{2}$ emission from warm gas and millimeter search for $^{12}$CO from cool
gas around 8 sun-like systems with ages $<$30 Myr measured upper limits 
$<$0.04 $M_{\rm Jup}$ gas within a few AU of the disk inner radius (1--40 AU) 
and $<$0.04 $M_{\earth}$ gas at 10--40 AU \citep{pas06}. \emph{Spitzer} IRS 
observations place 5$\sigma$ upper limits on the line flux from H$_{2}$ S(1) 
and S(0) toward HD 105, a G0V member of the Tucana-Horolgium association at 
40$\pc$ from the Sun, suggesting that less than 1500, 12, and 0.95$\Mearth$, 
respectively, at $T_{\rm gas}$\,=\,50, 100, or 200$\K$ remains in this disk at 
an age of 30\,Myr \citep{hol05}. Together, these observations suggest that warm
molecular hydrogen, with $T_{ex}$ = 100 K, dissipates from circumstellar disks 
on timescales $<$10 Myr.

\section{The Origin of the Atomic Gas}
The origin of the observed stable atomic gas in Keplerian rotation around 
$\beta$ Pictoris (described in the Introduction), like the micron-sized dust 
grains, is currently not well-constrained. There have been suggestions in the 
literature that the stable gas component is produced by infalling refractory
bodies \citep{bv07} or by collisions between dust grains \citep{cm07, fbw06}. 
We consider these possibilities and hypothesize that some of the observed 
atomic sodium is generated by photon-stimulated desorption from SiO$_{2}$-like 
surfaces. 

Infalling refractory bodies undoubtably contribute at least some portion of the
observed gas but whether all of the gas is generated by the evaporation of 
parent bodies is uncertain. The spatial distribution of the gas may provide a 
clue to its origin. The atomic gas is detected to distances of 150--200$\AU$ on
the southwest side and distances of 300--350$\AU$ on the northeast side of the 
disk. These distances correspond to the inner regions of the dusty disk that 
has been detected in thermal emission and scattered light to distances $>$1400 
AU. The SW/NE dust brightness asymmetry observed in scattered light is also 
seen in the resonantly scattered gas emission lines; however, the asymmetry in 
the \ion{Na}{1} emission is more pronounced. Finally, the gas and dust 
populations share an inner disk that is tilted $\sim$5$\arcdeg$ with respect 
to the outer disk \citep{bra04}. If infalling refractory bodies produce all of 
the observed gas, 
%(1) how does the gas migrate from regions near the star to distances of 
%    150--350$\AU$? 
%(2) 
why are the spatial distributions of the 
    gas and dust so similar?

To explain the similarity in the spatial distributions of the dust and gas 
around $\beta$ Pictoris, \cite{cm07} recently proposed that the atomic gas 
is produced via collisions between dust grains that vaporize at least some 
portion of the dust. Vaporization of dust grains in energetic collisions is 
believed to produce atomic gas in supernovae shocks and may generate a 
non-trivial fraction of the gas observed in the exospheres of Mercury and the 
moon \citep{kil01}. In the \cite{cm07} model, radiation pressure accelerates 
sub-blowout sized grains ($\beta$ meteoroids) to large radial velocities; these
small particles then collide with bound, orbiting grains with relative 
velocities as high as 90 km/sec (for carbonaceous grains). The observed 
abundances of the $\beta$ Pictoris dust and atomic gas may provide a clue to 
the origin of the stable atomic gas. If the gas is liberated in vaporizing 
collisions and no atomic species are selectively removed (e.g. via outgassing 
or differentiation), then the gas composition should be the same as the grain 
composition. Our models for the dust grains around $\beta$ Pictoris reproduce 
the \emph{Spitzer} IRS spectrum assuming a carbon relative to silicon abundance
ratio of $\simali$8.9, significantly smaller than the measured ratio of the 
column densities of atomic carbon and silicon gas, $\simali$500 \citep{rob06}. 

Photon-stimulated desorption (PSD) may produce a substantial fraction of the 
sodium observed in the tenuous exospheres of Mercury and the Moon 
\citep{leb06,lam03,ym99} and may produce some of the spatially-extended sodium 
gas observed via resonant scattering toward $\beta$ Pictoris. Since PSD 
produces gas directly from the dust, it might naturally explain the similarity 
in the gas and dust density distributions. The cross-section for PSD jumps 
dramatically at UV wavelengths, becoming most efficient at 
$\lambda < 2500\Angstrom$ \citep{ym99}. $\beta$ Pictoris has been observed 
extensively in the far and near ultra violet using \emph{FUSE} \citep{lde01, 
rob06}, \emph{HST} GHRS and STIS \citep{rob00}, with a continuum flux at 
$2000\Angstrom < \lambda < 2500 \Angstrom$ measured by \emph{IUE}, 
$F_{\lambda} > 4\times 10^{-11}\erg\s^{-1}\cm^{-2}\Angstrom^{-1}$ 
\citep{lhh95}. \cite{fbw06} model the observed stellar ultra violet spectrum of
$\beta$ Pic using a rotationally broadened 
($v_{\rm rot}$ = 130$\pm$4\,km\,sec$^{-1}$) 
PHOENIX model with effective temperature $T_\star\,=\,8000\,K$, 
and surface gravity, log\,$g$\,=\,4.2, 
suggesting a stellar UV photon production rate,
$L_{1000-2500 \Angstrom}$ = 2.0$\times$10$^{44}$ s$^{-1}$. 

The efficiency with which photons desorb sodium, averaged over the surface of
a sphere, is 
\begin{equation}
\epsilon = \frac{1}{4} Q \sigma
\end{equation}
\citep{ym99} where $Q$ is the photon-stimulated desorption cross section 
and $\sigma$ is the atomic surface coverage. Laboratory experiments 
measure $Q_{Na}$ $\sim$ (3$\pm$1) $\times$ 10$^{-20}$ cm$^{2}$ for lunar 
samples with temperatures of 250 K and using incident photons with $\lambda$ 
$<$ 2500 \AA, suggesting an efficiency of $\epsilon_{Na,250K}$ = 2.3 
$\times$ 10$^{-8}$ for materials with a lunar temperatures and compositions, 
$\sigma_{Na}$ = 3 $\times$ 10$^{12}$ cm$^{-2}$ \citep{ym99}. In addition, the 
photon-stimulated desoprtion efficiency of sodium is also function of substrate
temperature, $T_{\rm gr}$. We estimate 
\begin{equation}
\epsilon_{Na} = (6.93 \times 10^{-8}) 10^{-122.8 K /T_{\rm gr}}
\end{equation}
from measurements of sodium desorption yields in the laboratory \citep{ym04}, 
normalized to the sodium desorption efficiency at the moon as discussed above.

The rate at which atomic sodium is produced in the $\beta$ Pictoris disk is
\begin{equation}
\Gamma = \int_{R_{\rm min}}^{R_{\rm max}} 2 \pi r dr 
         \int_{z_{\rm min}}^{z_{\rm max}} dz\,
         \frac{L_{1000-2500 \Angstrom}}{4 \pi r^{2}}\, 
         \epsilon[T_{\rm dust}(r)]\,\pi a^{2}\,n_{\rm dust}(r,z)
\end{equation}
where $R_{\rm min} = 3\AU$ and $R_{\rm max} = 232\AU$ are the minimum and 
maximum radii, and $z_{\rm min} = -100\AU$ and $z_{\rm max} = 100\AU$ are the 
minimum and maximum heights at which the atomic sodium gas is observed. 
Since the \ion{Na}{1} gas in the $\beta$ Pic disk is located at distances of up
350 AU, well beyond our IRS slit, we use a dust density distribution inferred 
based on observations of the disk on large scales (rather than the dust
density profile that we derive in Section 3). Artymowicz (private 
communication) has inferred the quantity $\pi a^2 n_{\rm dust}$, summed over 
all grain sizes, from \emph{HST} STIS observations \citep{hea00} assuming a 
grain albedo 0.5:
\begin{equation}
\pi a^2 n_{\rm dust} = \frac{\tau_{\rm o}}{W} 
        \left[\left(\frac{r}{r_{\rm o}}\right)^{-4}+
        \left(\frac{r}{r_{\rm o}}\right)^{6} \right]^{-1/2} 
        \exp\left[-\left(\frac{z}{W}\right)^{0.7}\right]
\end{equation}
where $r_{\rm o}$\,=\,120\,AU, $W$\,=\,6.6\,$\left(r/r_{\rm o}\right)^{0.75}$, 
and $\tau_{\rm o}$\,=\,2$\times 10^{-3}$. We assume that the bulk of the
surface area is contained in 1 $\mu$m amorphous olivine grains and estimate
the grain temperature as a function of distance using Mie theory (the grains 
are spherical) and using laboratory-measured indices of refraction 
\citep{dor95}. If $R_\star = 1.69\,R_{\sun}$ and $T_\star = 8000\K$, we 
estimate a sodium production rate $1.3\times 10^{33}\s^{-1}$ from the grains 
via photon-stimulated desorption. We do not use the same dust spatial 
distribution as adopted in \S3 for modeling the IR emission because the 
desorption properties of porous aggregates invoked in \S3 are poorly known, 
while the desorption properties of compact grains (assumed in eq.[4])are better
quantified (e.g. experimental data are available). Moreover, the dust spatial 
distribution adopted in \S3 is already oversimplified. 

We estimate the number of sodium atoms in the $\beta$ Pictoris disk from 
observations of resonantly scattered \ion{Na}{1} and infer the time required to
produce the circumstellar sodium. Since the ionization potential of sodium is
5.1 eV and $\beta$ Pic possesses a high UV luminosity, the majority of the 
circumstellar sodium is expected to be ionized. \cite{fbw06} have written
a photoionization code that calculates the densities of neutral and ionized
elements from H to Ni, assuming that the $\beta$ Pic disk is optically thin
and has a solar composition; they estimate a sodium neutral fraction of 
3$\times$10$^{-4}$ at the disk midplane at 100$\AU$ from the central star. 
Since the sodium neutral fraction is a function of distance from the stellar
ionizing source, we have written a simple photoionization code which calculates
the densities of neutral and ionized carbon and sodium in the disk as a 
function of position assuming that (1) the gas is optically thin, (2) the disk 
gas contains only carbon and sodium, (3) ionization of carbon produces the 
electron density in the disk, (4) the gas temperature, 
$T_{\rm ex}$ = $T_{\rm gr}$, 
and (5) the circumstellar \ion{Na}{1} possesses a density distribution
\begin{equation}
n_{\rm NaI} = n_{\rm o} 
              \left[\left(\frac{r}{r_{\rm o}}\right)^{a}+
              \left(\frac{r}{r_{\rm o}}\right)^{b} \right]^{-1/2} 
              \exp\left[-\left(\frac{z}{\alpha r}\right)^{2}\right]
\end{equation}
\citep{bra04} where $n_{\rm o} = (1.02\pm0.04) \times 10^{-5}\cm^{-3}$, 
$r_{\rm o} = 117\pm3\AU$, $a = 0.94\pm0.06$, $b = 6.32\pm0.04$, and 
$\alpha = 0.168\pm0.05$. We use published carbon and sodium photoionization 
cross sections \citep{ver96} and radiative recombination coefficients 
\citep{vf96, nah95} to estimate the carbon and sodium densities assuming that
the atoms are in ionization equilibrium at each point in the disk. If the 
gas has a solar carbon:sodium abundance ratio, then the $\beta$ Pictoris disk 
possesses 4.6$\times$10$^{44}$ sodium atoms and a time of $\sim$13000 years is 
required to produce the sodium gas. Models of the breaking of atomic gas in
the $\beta$ Pic disk suggest that the lifetime of an atom in the disk is
$\sim$10$^{4}$ to 10$^{5}$ years \citep{fbw06}, suggesting that the sodium gas 
is in a steady state if it is produced by PSD and braked by Coulomb 
interactions.

We can estimate the radial dependence of the sodium number density and compare
it to that inferred from resonantly scattered sodium observations 
\citep{bra04}. If atomic sodium produced in the disk does not migrate radially,
then sodium gas generated by photon-stimulated desorption will have a number
density distribution
\begin{equation}
n(r,z) \propto \frac{L_{1000-2500 \Angstrom}}{4 \pi r^{2}}\, 
         \epsilon_{Na}[T_{\rm dust}(r,z)]\,\pi a^{2}\,n_{\rm dust}(r,z)
\end{equation}
where $r$ is the distance to the central star and $\pi a^2 n_{dust} $ is the
surface area contained in the dust grains. The predicted PSD-produced sodium 
density distribution at the disk midplane falls too quickly as a function of 
radius compared to that inferred from observations (see Figure 5). The rapid 
fall off in predicted sodium density is the result of the 1/$r^{2}$ dilution of
stellar photons. The sodium desorption efficiency is able to partially overcome
this effect because $\epsilon_{Na}[T_{dust}(r,z)]$ $\propto$ $r$ if the grains 
are 1 $\mu$m olivine spheres. However, even taking into account the temperature
dependence of the desorption efficiency, our predicted sodium number density 
falls off too quickly with radius by a factor of $r$. If PSD produces the bulk
of the observed sodium, sodium atoms must migrate to larger distances via 
radiation pressure.

%The observation that the scale height for \ion{Na}{1} is larger than that
%observed for the dust may be explained if the circumstellar sodium is produced
%via PSD. Laboratory studies suggest that sodium desorbs from lunar samples 
%with an escape velocity peaked at 800 m/s, corresponding to a temperature of 
%$\sim$900 K \citep{ym04}. This is the same temperature inferred for the 
%$\beta$ Pictoris gas based on measurements of the \ion{Na}{1} scale height at 
%100 AU \citep{fbw06}. Therefore, the kinetic energy transfered to desorbed 
%sodium atoms may heat the gas, producing the observed gas scale height. 
Our model for the origin of circumstellar sodium could be extended to other 
atomic species in the $\beta$ Pictoris disk. At the current time, few 
laboratory measurements have been made for the photon-stimulated desorption
rates of other atoms. Potassium is the only species, other than sodium, for 
which PSD rates have been measured. \cite{mad98} measure cross sections for
postassium desorption from Cr$_{2}$O$_{3}$ surfaces, $Q_{K}$ = 2 $\times$ 
10$^{-20}$ cm$^{2}$ at 2500 - 3500 \AA \ and 2 $\times$ 10$^{-19}$ cm$^{2}$ at 
1900 \AA. If photon-stimulated desorption of potassium has the same temperature
dependence as observed for sodium and the $\beta$ Pic circumstellar grains have
a solar potassium to sodium abundance ratio, $\sigma_{K}$ = 2.1 $\times$ 
10$^{11}$ cm$^{-2}$, then we estimate a potassium production rate in the
$\beta$ Pic disk of 3.3 $\times$ 10$^{32}$ K atoms s$^{-1}$, $\sim$3000 times 
less than the sodium production rate. Atomic potassium has not yet been 
detected in the $\beta$ Pictoris disk and 3$\sigma$ upper limits on the 
circumstellar potassium mass have not yet been reported in the literature.

\section{Discussion}
The simple dust model presented in \S3, 
consisting of cometary grains with a power-law size 
distribution and a modified power-law spatial distribution,
provides a reasonably good fit to the overall
SED and the observed {\it Spitzer} IRS spectrum. 
At a first glance, the model has many
free parameters: (1) $\amin$, $\amax$ and $\alpha$ for
the dust size distribution $dn/da$;
and (2) $\rmin$, $\rmax$, $f$, $\rp$, $\gamma$ for
the dust spatial distribution $dn/dr$.
With general constraints from the disk structure
and the dust absorption and emission properties,
we actually have only 2 free parameters $\alpha$ 
and $f$ left (see \S3 for details).
In so doing, 2 assumptions have been made:
(1) all grains have the same spatial distribution;
and (2) the grains producing the IRS emission spectrum
are also responsible for the optical scattered light
or their spatial distribution follows that of the dust
responsible for the optical scattered light.
Admittedly, the former assumption is oversimplified 
since the response of the grains to the stellar radiation 
pressure and Poynting-Robertson drag varies with grain size
and therefore the simple radial power law is probably
not all that representative of the actual grain distribution
uniformly with size. To justify the latter assumption,
a simultaneous modeling of both the scattered light and
IR emission is required. 
Nevertheless, the general conclusion for the dust
(the presence of crystalline silicates
and an inner warm dust component in the disk) 
remains valid.

We have presented a model for the production of circumstellar sodium gas
around $\beta$ Pictoris from dust via photon-stimulated desorption. Our model 
is able to generate the inferred circumstellar sodium within the residence time
of the gas but is unable to reproduce the radial dependence of the number 
density distribution if the sodium atoms do not migrate to larger distances via
radiation pressure. Our estimate for the time required to desorb the observed 
sodium is a function of (1) the circumstellar grain temperature, (2) the 
circumstellar gas temperature, and (3) the gas carbon:sodium abundance ratio. 

The photon-stimulated desorption rate of sodium from orbiting dust grains is 
dependent on the grain temperature with sodium more efficiently desorbed from 
warmer surfaces. Since the bulk of the surface area of the dust is contained in
the smallest grains, we estimated the grain temperature assuming that the 
grains have radii of $a$ = 1 $\mu$m, are composed of amorphous silicates, and 
are spherical. However, if the grains are very small (2$\pi a < \lambda$) then 
the grain temperature, $T_{\rm gr}$ = 
$\left[0.25 (R_\star/r)^{2}\right]^{0.2} T_\star$, 
and the sodium desorption rate,
$\Gamma$ = $3.7\times 10^{33}\s^{-1}$, and 4000 years are required to produce 
the sodium gas. Or, if the grains are very large (black bodies), then the grain
temperature, $T_{\rm gr}$ = 
$\left[0.25 (R_\star/r)\right]^{0.5} T_\star$, 
and the sodium desorption rate, 
$\Gamma$ = $2.5\times 10^{32}\s^{-1}$, and 58000 years are required to produce 
the sodium gas. 

In our simple model, we have assumed that the gas and dust have the same 
temperatures and we have explored the possible range in dust temperature 
profiles. We also consider how changes in the gas temperature profile affect
our estimates for the sodium neutral fraction and therefore the total number
of sodium atoms in the disk. The estimated neutral fraction of sodium is weakly
dependent on the gas temperature and composition. The radiative recombination 
coefficients for carbon and sodium decrease by factors of $\sim$5 if the gas 
temperature increases from 10 K to 100 K or 100 K to 1000 K. If the gas has a 
temperature profile similar to small grains or black bodies (instead of 1 
$\mu$m olivine spheres), then we estimate that the $\beta$ Pictoris disk 
possesses either 7.0$\times$10$^{44}$ or 5.4$\times$10$^{44}$ sodium atoms, 
respectively, similar to the 4.6$\times$10$^{44}$ sodium atoms expected if the 
gas has the temperature profile expected from 1 $\mu$m amorphous silicate 
grains. The assumed abundance of sodium relative to carbon also affects the 
total number of sodium atoms inferred from the neutral sodium observations. In 
our model, we have assumed that the gas has a solar composition; however, a 
recent inventory of the circumstellar atomic gas around $\beta$ Pictoris 
suggests that carbon is enriched compared to iron and oxygen by factors of 16 
and 18, respectively \citep{rob06}. If the $\beta$ Pictoris disk is enriched in
carbon relative to sodium by a factor of 16, then the disk possesses 
1.7$\times$10$^{44}$ sodium atoms, somewhat less than the number inferred if 
the gas has a solar composition.

%\newpage
%\begin{equation}
%T_{gr} =  \left[ \frac{1}{4} \left(\frac{R_{*}}{D} \right)^{2} T_{*}^{5}
% \right]^{1/5}
%\end{equation
%\newpage

\section{Conclusions}
We have obtained \emph{Spitzer Space Telescope} IRS spectra 
of the 12\,Myr old debris disk around $\beta$ Pictoris. 
We find that:

%1. The \cite{lg98} model for the composition 
%and spatial distribution of dust 
%around $\beta$ Pictoris reproduces 
%the observed spectrum well. However, fluffy 
%cometary and crystalline olivine aggregates 
%alone are not sufficient to explain
%the excess emission observed at $\lambda$ $<$ 15 $\mu$m. 
%An additional warm amorphous silicate grain population 
%is needed to account for the short wavelength excess emission

1. In addition to the 10$\mum$ silicate emission feature 
   originally detected in ground-based observations, 
   we, for the first time, also observe weak crystalline 
   silicate emission features at 28$\mum$ and 33.5$\mum$.

2. The IRS dust emission spectrum and the overall SED
   are well reproduced by a dust model consisting of 
   fluffy cometary and crystalline olivine aggregates.

3. No H$_{2}$, [\ion{S}{1}], \ion{Fe}{2}, or \ion{Si}{2} 
   emission is detected. Our 3$\sigma$ upper limits 
   suggest that $<$17$\Mearth$ H$_{2}$ remains in 
   the disk significantly less than the previously reported detections 
   of H$_{2}$ S(1) and S(0) emission. 
   The circumstellar disk around $\beta$ Pictoris has 
   too little gas to support the formation of giant planets. 

4. Some of the observed resonantly scattered \ion{Na}{1}, 
   observed at visual wavelengths, may be produced 
   via photon-stimulated desorption; the timescale to generate 
   all of the inferred circumstellar sodium is 13000 yr, 
   approximately the residence time of the gas in the disk if 
   the gas is braked by Coulomb forces.

\acknowledgements
We would like to thank A. Brandeker for providing the rotationally broadened
PHOENIX stellar atmosphere model for $\beta$ Pictoris and I. Mann for providing
us a preprint of their manuscript on their model for a collisional origin for
the $\beta$ Pic atomic gas prior to publication. We would also like to thank
our two anonymous referees and E. Chiang, D. Hollenbach, M. Jura, J. Najita, 
A. Roberge, N. Samarasinha, W. Sherry, A. Weinberger, K. Willacy, and Y. Wu for
their helpful comments and suggestions. Support for this work at NOAO was 
provided by NASA through the Spitzer Space Telescope Fellowship Program, 
through a contract issued by the Jet Propulsion Laboratory, California 
Institute of Technology under a contract with NASA. Support for this work at 
the University of Missouri was provided in part by the University of Missouri 
Summer Research Fellowship, the University of Missouri Research Board, a 
NASA/HST Theory Program grant, and a NASA/Spitzer Theory Program grant. Support
for this work at the University of Arizona was provided by NASA through 
Contract Number 1255094 issued by JPL/Caltech.

\clearpage
\begin{deluxetable}{lccccc}
\tabletypesize{\small}
\tablecaption{Spitzer IRS Line Flux Upper Limits for $\beta$ Pictoris}
\tablehead{
    \omit &
    \omit &
    \omit &
    \colhead{50$\K$} &
    \colhead{100$\K$} &
    \colhead{Atomic/Molecular} \\
    \colhead{Gas Species} &
    \colhead{Wavelength} &
    \colhead{Line Flux} & 
    \colhead{M$_{\rm gas}$} &
    \colhead{M$_{\rm gas}$} &
    \colhead{Data} \\
    \omit &
    \colhead{($\mu$m)} &
    \colhead{($\erg\s^{-1}\cm^{-2}$)} &
    \colhead{($\Mearth$)} &
    \colhead{($\Mearth$)} &
    \colhead{References} \\
}
\tablewidth{0pt}
\tablecolumns{3}
\startdata
H$_{2}$ S(2)     & 12.279 & $<$5.0$\times$10$^{-14}$ & $<$1.6$\times$10$^{11}$ 
                 & $<$1.7$\times$10$^{4}$ & 2,4 \\
H$_{2}$ S(1)     & 17.035 & $<$1.2$\times$10$^{-14}$ & $<$2.1$\times$10$^{5}$ 
                 & $<$17 & 2,4 \\
$[$\ion{S}{1}$]$ & 25.249 & $<$8.6$\times$10$^{-14}$ & $<$0.047 & $<$0.00016 
                 & 1 \\
\ion{Fe}{2}      & 25.988 & $<$6.1$\times$10$^{-14}$ & $<$0.29 & $<$0.0011 
                 & 3 \\
H$_{2}$ S(0)     & 28.221 & $<$4.1$\times$10$^{-14}$ & $<$3400 & $<$43 
                 & 2,4 \\
\ion{Si}{2}      & 34.814 & $<$4.9$\times$10$^{-13}$ & $<$0.028 & $<$0.00046 
                 & 1 \\
\enddata
\tablerefs{(1) Haas, Hollenbach, \& Erickson 1991;
  (2) Jennings, Weber \& Brault 1987; 
  (3) Nussbaumer \& Storey 1980;
  (4) Wolniewicz, Simbotin, \& Dalgarno 1998}
\end{deluxetable}

\clearpage

\begin{figure}
\figurenum{5}
\plotone{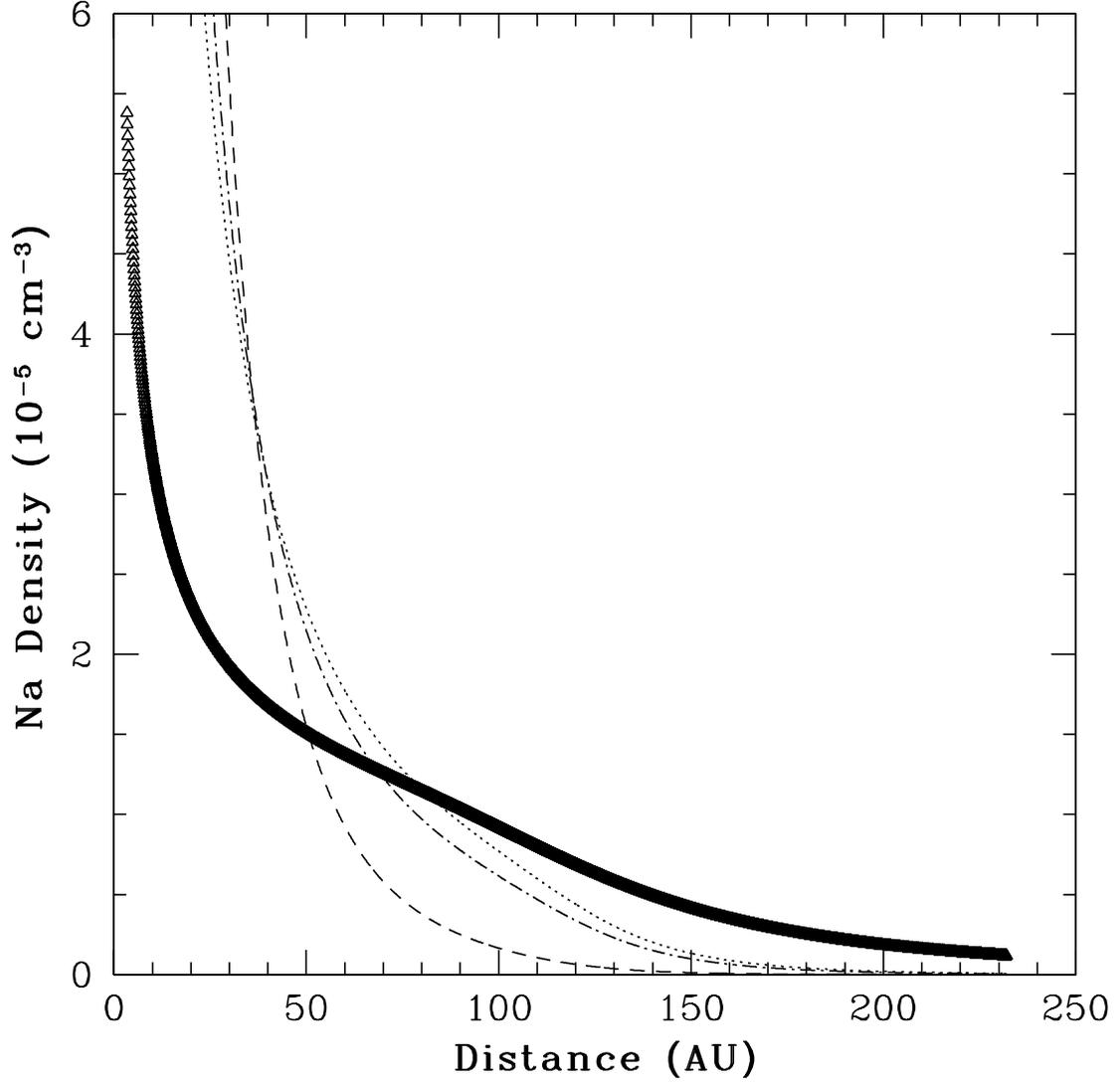}
\caption{
The estimated number density distribution of \ion{Na}{1} atoms at the midplane
of the $\beta$ Pic disk, inferred from observations of resonantly scattered 
\ion{Na}{1} \citep{bra04}, is shown with triangles. For comparison, we overplot
the number density distributions of Na gas produced via PSD, assuming that the 
grains are small (2$\pi a$ $<$ $\lambda$; dotted line), large (black bodies; 
dashed line), or 1 $\mu$m olivine spheres (dashed-dotted line). Our models are 
normalized such that they contain the same mass as the fitted model.}
\end{figure}

\end{document}